\documentclass[12pt]{article}

\usepackage{cite,amstext,amsfonts,amsmath}
\usepackage{geometry,color}

\def\del          {\partial}

\def\ii           {{\rm i}}
\def\d{\text{d}}

\def\cald         {{\cal D}}
\def\cale         {{\cal E}}
\def\calf         {{\cal F}}
\def\calg         {{\cal G}}

\def\call         {{\cal L}}
\def\calm         {{\cal M}}
\def\caln         {{\cal N}}

\def\calr         {{\cal R}}

\def\calw         {{\cal W}}
\def\calz         {{\cal Z}}

\def\be{\begin{equation}}
\def\ee{\end{equation}}
\def\bseq{\begin{subequations}}
	\def\eseq{\end{subequations}}

\def\bea{\begin{eqnarray}}
\def\eea{\end{eqnarray}}

\def\bseq{\begin{subequations}}
	\def\eseq{\end{subequations}}

\title{Three-forms, Supersymmetry\\ and String Compactifications}
\date{}
\author{Fotis Farakos$^{a}$, Stefano Lanza$^{b}$, Luca Martucci$^{b}$ and Dmitri Sorokin$^{b}$ 
}

\begin{document}

\maketitle

\begin{center}

\vspace{-0.6cm}
    \textit{\small ${\, }^{(a)}$KU Leuven, Institute for Theoretical Physics,\\ Celestijnenlaan 200D, B-3001 Leuven, Belgium } \\ 
	\vspace{0.2cm}
	\textit{\small ${\, }^{(b)}$Dipartimento di Fisica e Astronomia ``Galileo Galilei",  Universit\`a degli Studi di Padova \\ 
		\& I.N.F.N. Sezione di Padova, Via F. Marzolo 8, 35131 Padova, Italy}
\end{center}

\abstract{\noindent
	We review a duality procedure that relates standard matter-coupled $\caln=1$ supergravity to dual formulations in which auxiliary fields are replaced by field-strengths of gauge three-forms. As examples, we consider the dualization of the rigid Polonyi model and of effective field theories associated with 
	Type IIA string compactifications with fluxes in supergravity.}

\section{Introduction}

In four-dimensions the gauge three-forms  do not carry any propagating degree of freedom. Nevertheless, their presence can have non-trivial dynamical consequences. In particular, they can play an important role in supergravity and string theory scenarios  (see e.g. \cite{Ovrut:1997ur,Bousso:2000xa,Feng:2000if,Groh:2012tf,Bielleman:2015ina,Farakos:2016hly,Buchbinder:2017vnb} and references therein).

In generic string compactifications to four-dimensions, gauge three-forms naturally arise from the KK reduction of higher dimensional gauge forms. The on-shell value of the corresponding field-strengths is dual to the value of the internal fluxes threading the compactification space. 
The problem of formulating the low-energy effective theory of string flux compactifications in terms of gauge three-forms, rather than in terms of the dual internal fluxes,  was recently addressed in \cite{Bielleman:2015ina}, focusing on the bosonic sector. These kinds of effective theories should admit a supersymmetric completion, but the effective theories  obtained in  \cite{Bielleman:2015ina} did not fit into any of the previously known supersymmetric models including gauge  three-forms \cite{Stelle:1978ye,Ogievetsky:1978mt,Ogievetsky:1980qp,Gates:1980ay,Gates:1980az}.

In this contribution we review the results of \cite{Farakos:2017jme} in which  a new broad family of rigid and local  $\mathcal N=1$ supersymmetric models including gauge three-forms  was proposed. The derivation of these models is based on a novel  non-linear duality between conventional chiral and three-form multiplets, which has the effect of promoting to dynamical variables  part of the coupling constants defining the superpotential for the conventional chiral multiplets. This duality procedure provides a  four-dimensional supersymmetric realization of what expected from string flux compactifications and the generalization thereof. Indeed, as reviewed below,  these general results allow for  a supersymmetric effective description of IIA compactifications with R-R fluxes in terms of gauge three-forms.

\section{Rigid Polonyi model} 
\label{Polonyi}

Let us first demonstrate how the dualization procedure works in an example describing the dynamics of a single chiral multiplet (in the conventions of \cite{Wess:1992cp}) 
\be\label{Phi}
\Phi =\varphi +\sqrt{2}\theta^\alpha \psi_\alpha +\theta^2 f \, , \qquad \bar D_{\dot\alpha}\Phi=0, 
\ee
which undergoes spontaneous supersymmetry breaking.

The $\mathcal N=1$ superspace Lagrangian of the Polonyi model in the rigid limit is 
\begin{equation}\label{Polonyi_A}
\call= \int \d^4\theta\, \Phi \bar{\Phi} + \left( \int \d^2\theta\, b\,\Phi + c.c. \right) \, , 
\end{equation}
where $b$ is a complex constant. For the component fields  we find 
\be\label{Polonyi_CompA}
\call= - \del_m \bar{\varphi} \del^m \varphi - \ii \psi \sigma^m \del_m \bar{\psi}+ \bar{f}f+ b f+\bar{b} \bar{f}\,.
\ee
Once the auxiliary field $f$  takes its on-shell value $f = - \bar b$, the Lagrangian \eqref{Polonyi_CompA} becomes 
\be\label{Polonyi_CompB}
\call= - \del_m \bar{\varphi} \del^m \varphi - \ii \psi \sigma^m \del_m \bar{\psi}-{b}\bar{b}\,.
\ee
We see that the last term in \eqref{Polonyi_CompB}, associated with the on-shell value of $f$, 
contributes to the vacuum energy. 
Supersymmetry here is spontaneously broken and $\psi_\alpha$ becomes a Goldstone fermion. 

Our goal is to find a way to generate the constant $b$ \emph{dynamically} without adding a superpotential. 
This can be achieved by trading the auxiliary field of $\Phi$ for the field-strength $F_4\equiv \d C_3$ of a complex gauge three-form $C_{3}$. Let us call $S$ the special chiral multiplet
\be\label{S}
S =\varphi +\sqrt{2}\theta^\alpha \psi_\alpha +\theta^2 f_S \, ,
\ee
whose highest component $f_S$ has the following constrained form 
\be\label{deffS}
f_S = {}^*{F_4} = \del_m C^m\,, \quad\text{with}\quad C^{m} = \frac{1}{3!} \varepsilon^{mnpq} C_{npq}\,.
\ee
The free Lagrangian for $S$, which up to boundary terms is described in superspace by $\tilde \call=\int \d^4 \theta\, S \bar S$, has the following component form 
\be\label{Polonyi_Comp_DualA}
\tilde\call = - \del_m \bar{\varphi} \del^m \varphi - \ii \psi \sigma^m \del_m \bar{\psi}+ |{}^*{F_4}|^2 -\del_m (C^m\, {}^*{\bar F_4})-\del_m(\bar{C}^m \, {}^*{F_4})\,,
\ee
where the total derivative terms have been added to ensure the correct variation of the gauge three-form. 
Indeed, varying with respect to $C^m$ and imposing the gauge invariant boundary conditions $\delta F_4|_\textrm{bd} = 0$, we get
\be
\del_m {}^*{F_4} =0 \quad \Rightarrow \quad {}^*{F_4}= b \, , 
\ee
which leads again to \eqref{Polonyi_CompB}, with the important difference that now $b$ appears as a dynamical integration constant parameter. 

Having found the dual model, a further question one can address is whether it is possible to pass from \eqref{Polonyi_A} to \eqref{Polonyi_Comp_DualA}  in a manifestly supersymmetric way. To this end, let us note that the special chiral superfield $S$ can be parametrized as  follows \cite{Gates:1980az}
\be\label{superS}
S= -\frac{1}{4} \bar{D}^2 \bar{\Sigma} \, , 
\ee
where $\Sigma$ is a complex linear multiplet  subject to the superspace constraint $\bar{D}^2 \Sigma = 0$. 
The gauge three-form resides in the component field 
\be\label{sS}
\frac{1}{2} \bar{\sigma}^{m\,\dot\alpha \alpha}
\left[D_\alpha,\bar{D}_{\dot{\alpha}}\right] \Sigma| = C^{m}(x).
\ee
Due to the gauge invariance of $S$ under $\Sigma \rightarrow \Sigma + L$  (where $L$ is a real linear superfield parameter), the chiral multiplet $S$ contains only the gauge-invariant field strength of $C_3$ as in \eqref{S}. Note also that the complex linear superfield can be expressed in terms of a generic Weyl spinor superfield $\Psi_\alpha$ as $\bar{\Sigma} =D^\alpha \Psi_\alpha$, which can be used to derive the equations of motion of $\Sigma$. 

We are now ready to show how to get the new formulation from the old one by a manifestly supersymmetric duality procedure. Let us promote the complex constant $b$ appearing in \eqref{Polonyi_CompA} to a chiral multiplet $X$ and add to \eqref{Polonyi_A} a term which contains the complex linear multiplet $\Sigma$ 
\begin{equation}\label{Polonyi_Dual}
\begin{split}
\call''=& \int \d^4\theta\, \Phi \bar{\Phi} + \left(\int \d^2\theta X\Phi + c.c. \right) - \left[ \int \d^2 \theta \left(-\frac14 \bar{D}^2 \right) \left(\bar{X} \Sigma \right) + c.c \right].
\end{split}
\end{equation}
By varying \eqref{Polonyi_Dual} with respect to $\Sigma$ we get $X=b$ and hence recover
\eqref{Polonyi_A}. To find the dual formulation we vary \eqref{Polonyi_Dual} with respect to $X$ and $\Phi$, 
which produces 
\begin{align}
	&\delta X:\; \Phi = -\frac{1}{4} \bar{D}^2 \bar{\Sigma} = S, \qquad \delta \Phi:\; X = \frac{1}{4} \bar{D}^2 \bar{S} \label{Polonyi_deltaPhi}.
\end{align}
Substituting \eqref{Polonyi_deltaPhi} back into \eqref{Polonyi_Dual} we get the dual Lagrangian
\begin{equation}\label{Polonyi_Dual_Fin}
\begin{split}
\call'=& \int \d^4\theta\, S \bar{S} +\call_{\rm{bd}}
\end{split}
\end{equation}
with 
\be\label{Polonyi_Dual_Bd}
\call_{\rm{bd}} = \left(\int \d^2\theta X\Phi + \frac14 \int\d^2\theta\bar D^2 
\left(\bar{X} \Sigma\right) 
\right) + c.c.  
\ee
where $X$ and $\Phi$ take the values \eqref{Polonyi_deltaPhi}.

One can check  that the component form of \eqref{Polonyi_Dual_Fin} is given by \eqref{Polonyi_Comp_DualA}, while \eqref{Polonyi_Dual_Bd} is the superfield extension of the boundary term appearing in \eqref{Polonyi_Comp_DualA}. Note that this is directly produced by the dualization procedure.


\section{Type IIA effective field theory}

The dualization procedure of the previous example can be extended to more general globally and locally supersymmetric theories.  For instance, let us consider a rigid theory  with a set of  chiral superfields $\Phi^A$ and a superpotential of the form
\be\label{Sup}
W=e_A\Phi^A+m^A\calg_{AB}(\Phi)\Phi^B+\hat W(\Phi) \, ,
\ee
where $e_A$ and $m^A$ are real constants and $\hat W(\Phi)$, $\calg_{AB}(\Phi)$ are arbitrary holomorphic functions which may also depend on additional chiral superfields.  In \cite{Farakos:2017jme}, it was shown that such a theory admits a dual formulation in which the auxiliary fields $f^A$ of the chiral multiplets (along with the ones of the supergravity multiplets) and eventually the constants $e_A$ and $m^A$ get replaced by combinations of field-strengths $F^A_4=\d C^A_3,\tilde F_{4A}=\d \tilde C_{3A}$ associated with pairs of gauge three-forms $C^A_3,\tilde C_{3A}$. The resulting multiplets were dubbed double three-form multiplets.

Now,  being non propagating, the three-forms can be integrated out by means of their equations of motion. As a result, the parameters $(e_A, m^A)$ appearing in \eqref{Sup} are generated dynamically as expectation values of the four-form field strengths, as in the simple example of the previous section.  In turn, this implies that the form of the potential of the scalar fields, governed by the superpotential \eqref{Sup} in the original formulation, is now determined by the underlying four-forms. 

By starting from a super-Weyl invariant superspace formulation of supergravity, the same procedure can be applied to locally supersymmetric theories as well \cite{Farakos:2017jme}. The only difference is that the chiral fields  dualized to double three-form multiplets now include the conformal compensator.   In the following, we will review the main points of  this dualization procedure. For the sake of concreteness, we will focus on the particular example provided by the effective theories  of type IIA flux compactifications, whose standard formulation will be reviewed in subsection \ref{sec:IIAreview}. 
The derivation of the dual formulation will then be presented in subsection \ref{sec:sugraduality} and can be easily adapted to more general models \cite{Farakos:2017jme}.

\subsection{Effective $\caln=1$, $D=4$ theories from type IIA}
\label{sec:IIAreview}

The four-dimensional theory that we are going to examine is the $\caln=1$ supergravity arising from the compactifications of type IIA string theory on a Calabi-Yau three-fold $Y$ with O6-planes, studied, for example, in \cite{Grimm:2004ua}. 

The gauge sector of the ten-dimensional type IIA effective theory consists of the $p$-forms $A_p$ (with $p=2n-1$). Their  $(p+1)$-form field strengths $G_{p+1}$ can be compactly arranged into the polyform
\be
\label{Polyform}
\mathbf{G} = (\mathbf{F}+\overline{\mathbf{G}}) \wedge e^{B_2},
\ee
where $\mathbf{F} \equiv \d \mathbf{A}$ and $\overline{\mathbf{G}}$ is the polyform of the internal  fluxes (that is, those with `legs' along the Calabi-Yau space only). The higher-rank forms are related to the lower-rank ones by the ten-dimensional Hodge duality $G_{2n}=*_{10} G_{10-2n}$.

The internal flux quanta $e_0,e_i,m^i$ and $m^0$ are defined as follows
\be\label{Flux_quanta}
\begin{aligned}
	m^0 \equiv \overline{G}_0, \quad m^i \equiv \int_Y \tilde{\omega}_{4}^i \wedge \overline{G}_2, \quad e_i = \int_Y \omega_{2i} \wedge \overline{G}_4,\quad e_0 \equiv \int_{Y}\overline{G}_6 , 
\end{aligned}
\ee
 where $\omega_{2i}$ and $\tilde{\omega}_4^i$ are harmonic bases of the CY orientifold-odd $H_-^2(Y, \mathbb{Z})$ and orientifold-even $H_+^4(Y,\mathbb{Z})$ (with $i=1,\ldots, h^{1,1}_-(Y)$ ), respectively. 
 
 Expanding the field strengths in the external (that is, four-dimensional space-time) and the internal parts, we may write the fluxes as expansions over the internal bases
\begin{align}
\nonumber
    &G_0 = m^0,&\quad &G_2 = m^i\, \omega_{2i}+\ldots,\\\label{FSExp}
	&G_4 = F_4^0+ e_i\, \tilde{\omega}_4^i+ \ldots, &\quad &G_6 = F_4^i \wedge \omega_{2i} + e_0 \omega_6+\ldots,
	\\\nonumber
	&G_8 = \tilde{F}_{4i} \wedge \tilde{\omega}_4^i + \ldots, &\quad &G_{10} = \tilde{F}_{40} \wedge \omega_6 + \dots\, , 
\end{align}
up to other contributions which we are  not interested in. Here $e_0,e_i,m^i$ and $m^0$ are the flux quanta defined in \eqref{Flux_quanta} and $(F_4^0,F_4^i,\tilde{F}_{4i},\tilde{F}_{40})$ are field-strengths in the external $4D$ space-time which are related by the $10D$ Hodge duality to the values of the internal fluxes.

Let us now consider the scalar sector. We shall only focus on the closed string moduli. One set of these moduli originates from the expansion of the CY K\"ahler form $J$ and the NS-NS two-form $B_2$ in the basis of orientifold-odd integral harmonic 2-forms $\omega_{2i}$
\be
J=v^i\omega_{2i}\,,\qquad B_2=b^i\omega_{2i}.
\ee
The moduli $v^i$ and $b^i$, along with their supersymmetric partners, combine into $n= h^{1,1}_-(Y)$ $4D$ chiral multiplet $\Phi^i$ whose lowest components are
\be \label{Moduli_Comp}
\Phi^i| \equiv \Phi^i|_{\theta=\bar\theta=0}= \varphi^i=v^i-\ii b^i \, . 
\ee
Another set of moduli is given by the complex structure, the dilaton and the internal R-R three-form moduli which combine into additional chiral multiplets $T^q$, with  $q=1,\ldots, h^{2,1}(Y)+1$. In the following, we will denote $t^q\equiv T^q|$ and $F^q_T\equiv-\frac{1}{4}D^2 T^q |$.

In the large volume and constant warping approximation, the K\"ahler potential can be split into two contributions as
\be\label{EFT_K}
\mathcal K(\Phi,\bar\Phi,T,\bar T)=K(\Phi,\bar\Phi)+ \hat K(T,\bar T)\,,
\ee
where the K\"ahler potentials $K(\Phi,\bar\Phi)$ and $\hat{K}(T,\bar T)$ satisfy the no-scale conditions
\be\label{eq:EFT_NS}
K^{\bar\jmath  i}K_{i}K_{\bar\jmath}=3, \quad \hat  K^{\bar rq} \hat  K_q \hat  K_{\bar r}=4\,,
\ee
where $K_{i}\equiv \frac {\partial K}{\partial \Phi^i}$, $ K_{i\bar\jmath}\equiv\frac {\partial^2  K}{\partial  \Phi^i\partial \bar\Phi^{\bar\jmath}}$, \ldots and $K^{ \bar\jmath i}$ is the inverse of the K\"ahler metric $K_{ i\bar\jmath}$, and similarly for $\hat K$.
\if{}
$ \hat  K_q\equiv\frac {\partial \hat  K}{\partial T^q}$, $ \hat  K_{q\bar r}\equiv\frac {\partial^2  \hat  K}{\partial T^q\partial \bar T^{\bar q}}$, \ldots, and  $ \hat  K^{\bar rq}$ is the inverse of the K\"ahler metric $ \hat  K_{q\bar r}$.
\fi
We assume that the K\"ahler potential $K(\Phi,\bar\Phi)$ solely depends on the real parts of the superfields $\Phi^i$ as follows
\be\label{eq:EFT_K_Phi}
K(\Phi,\bar\Phi)=-\log\left[\frac1{3!}k_{ijk}(\textrm{Re}\,\Phi^i)( \textrm{Re}\,\Phi^j)(\textrm{Re}\,\Phi^k)\right] \, , 
\ee
where $k_{ijk}$ are the triple intersection numbers $ k_{ijk}=\int_Y \omega_{2i}\wedge\omega_{2j}\wedge\omega_{2k}$.

The last ingredient which defines the theory is the flux-induced superpotential, which depends only on the chiral multiplets $\Phi^i$ (see also \cite{Bielleman:2015ina,Carta:2016ynn}) 
\be\label{EFT_Superp}
W=e_0+\ii e_i\Phi^i-\frac 12 k_{ijk}m^i\Phi^j\Phi^k+\frac \ii6 m^0k_{ijk}\Phi^i\Phi^j\Phi^k.
\ee
As we will see, \eqref{EFT_Superp} is a particular case of the locally supersymmetric counterpart of \eqref{Sup}, where the flux quanta \eqref{Flux_quanta} appear.

\subsection{The dual formulation}
\label{sec:sugraduality}

The expansion \eqref{FSExp} of the ten-dimensional field-strengths produces external space-time field-strengths and the four-dimensional effective field theory is naturally endowed with non-propagating gauge three-forms. Therefore, as in the example of Section \ref{Polonyi}, we aim at making the $2n+2$ constants $(e_{A},m^{A})$ in \eqref{EFT_Superp} dynamical by replacing them with their dual four-dimensional four-form field-strengths given in \eqref{FSExp}. 

The starting point is the super-Weyl invariant supergravity theory coupled to $n+1$ chiral multiplets $\calz^A=(\calz^0,\calz^i)$, among which we single out the chiral compensator $Z$ and the chiral matter superfields as follows
\be
\calz^0\equiv Z\,,\quad\calz^i\equiv \ii Z \Phi^i \,.
\ee  
In addition, there are also the `spectator' chiral multiplets $T^q$ which play an important role in determining the final structure of the Lagrangian for the four-forms in the dual formulation.

Introducing the `kinetic potential' $\Omega({\mathcal Z},\bar {\mathcal Z},T,\bar T)= |Z|^{\frac 23} \exp \left( - \frac{1}{3} {\mathcal K}\right) $ and the homogeneous superpotential $\calw(\calz,T)$  with the following homogeneity properties under the Weyl rescaling
\be\label{eq:Memb_SW_Hom}
\Omega(\lambda \mathcal Z,\bar\lambda\bar {\mathcal Z},T,\bar T)=|\lambda|^{\frac 23}\Omega({\mathcal Z},\bar {\mathcal Z},T,\bar T)\,,\qquad 
\calw(\lambda \mathcal Z,T)=\lambda\calw(\mathcal Z,T) , 
\ee
we construct the super-Weyl invariant superfield Lagrangian for chiral matter coupled to old minimal supergravity which takes the form \cite{Kugo:1982mr} 
\be\label{Sugra_Lag}
\call=-3\int \d^4\theta\, E\,\Omega({\mathcal Z},\bar {\mathcal Z},T,\bar T) 
+ \left(\int\d^2\Theta\,2\cale\,\calw(\calz, T) + c.c \right) \, . 
\ee 
Here $E$ is the Berezin super-determinant and $\int\d^2\Theta\,2\cale$ is the chiral superspace measure which transform under the Weyl rescaling as follows
\be
E \,\,\rightarrow\,\, |\lambda|^{-\frac 23} E,\qquad \d^2\Theta\,2\cale \,\,\rightarrow \,\, \lambda^{-1} \d^2\Theta\,2\cale.
\ee
We will focus on the specific class of the K\"ahler potentials \eqref{EFT_K}, \eqref{eq:EFT_NS}, \eqref{eq:EFT_K_Phi} and the homogeneous superpotential $\calw$ corresponding to the standard superpotential 
\eqref{EFT_Superp}. This can be written in the form 
\be\label{Sugra_SupPot}
\calw(\calz)\equiv e_A\calz^A+m^A \calg_{AB}(\calz)\calz^B \, , 
\ee
with $\calg_{AB}(\calz)\equiv \partial_A\partial_B \calg(\mathcal Z)$, where
\be\label{Sugra_Pot}
\calg(\calz)=\frac 1{6\calz^0} k_{ijk}\calz^i\calz^j\calz^k \,.
\ee 
Notice that $\calg$ is homogeneous of degree-two $\calg(\lambda \calz) = \lambda^2 \calg(\calz)$, consistently with the degree-one homogeneity of $\calw$. We may regard $\calg$ as a prepotential defining a special K\"ahler geometry locally parametrized by the homogeneous coordinates ${\mathcal Z}^{A}$.

As explained in \cite{Farakos:2017jme},  one can perform a dualization to a new theory in which the chiral multiplets $\mathcal Z^{ A}$ are replaced by special chiral fields $S^A$. However, in this more general case, the linear relation \eqref{superS} is substituted by
\be\label{mZA} 
S^{ A}= \frac14(\bar\cald^2-8\calr)\left[\calm^{ {AB}}(\Sigma_{ B}-\bar\Sigma_{ B})\right] \, , 
\ee 
with $\calm_{AB} \equiv \rm{Im}\, \calg_{AB}$ and $\calm^{AB} \equiv (\calm_{AB})^{-1}$. Here $\Sigma_A$ are complex linear multiplets (i.e.\ $-\frac14(\bar\cald^2-8\calr)\bar\Sigma_A=0$) which contain the double sets of real gauge three-forms $(A_3^{A},\tilde A_{3A})$ among their componets	
\be
	\frac12 \bar\sigma_m^{\dot \alpha \alpha} [{\cal D}_\alpha , \bar{\cal D}_{\dot \alpha} ] \Sigma_A |
    = - ({}^*\!\tilde A_{3A} + 
    {\cal G}_{AB} \, {}^*\!A_3^{B})_m  \, . 
	\ee
The special chiral multiplets $S^A$ parametrize the gauge invariant degrees of freedom of the double three-form multiplets $\Sigma_A$.

In \cite{Farakos:2017jme}, to which we address the reader for details about the dualization procedure, it was shown that \eqref{Sugra_Lag} can be dualized to the Lagrangian
\be\label{Sugra_Lag_Dual}
\call_{\textrm{dual}}=-3\int\d^4\theta\,E\,\Omega(S,\bar {S},T,\bar T)+{\mathcal L}_{\rm bd} \, .
\ee
 In practice, the dualization has replaced  $\calz^A$ with $S^A$ and removed the superpotential from \eqref{Sugra_Lag}, coherently with the presence of four-forms. Furthermore boundary terms are produced, which are needed to ensure the correct variation of the action.

In order to arrive at a more standard Einstein-frame formulation, one should fix the super-Weyl invariance, for instance by setting $S^0=1$. If  we focus on the purely bosonic sector and ignore fermions, this immediately gives $s^0=1$ and $F^0_{S}=0$. From the component expansion of \eqref{mZA} one can then extract the following relations 
\be\label{Sugra_M+F}
\begin{aligned}
	\bar M &=-  
	\frac{\ii}{2} {\cal M}^{0B}(z,\bar z)\left[ \,{}^*\!\tilde F_{4B}
	+\bar{\cal G}_{BC}(\bar z) \,{}^*\!F^C_{4}\right] \, ,\\
	F^{i}_S &= \bar M z^{i}  
	+ \frac{\ii}{2} {\cal M}^{iB}(z,\bar z) \left[\,{}^*\!\tilde F_{4B}
	+  \bar{\cal G}_{BC}(\bar z) \,{}^*\!F^C_{4}\right] \, ,
\end{aligned}
\ee
where $F_4^A\equiv\d A^A_3$, $\tilde F_{4A}\equiv\d \tilde A_{3A}$ and $M$ is the complex scalar auxiliary field of the gravity multiplet.
These equations explicitly show that $M$   and the $\Theta^2$-components of $S^i$ are expressed in terms of the four-form field strengths, which is the core of our dualization procedure.

Finally, after performing a standard Weyl rescaling in order to go to the Einstein frame and integrating out the auxiliary fields $F^r_T$ of the `spectators' $T^r$, one derives from \eqref{Sugra_Lag_Dual} the following Lagrangian for the bosonic fields
\be\label{Sugra_bos}
e^{-1} {\cal L}_{\rm bos} =  -\frac{1}{2} R 
- K_{i\bar\jmath} (\varphi,\bar\varphi)\,\partial \varphi^i \partial \bar \varphi^{\bar\jmath} - \hat K_{q \bar r} (t,\bar t)\,\partial t^q \partial \bar t^{\bar r} + e^{-1}\,{\cal L}_{\text{3-form}} + e^{-1}\,{\mathcal L}_{\rm bd}\, , 
\ee
where $\varphi^i$ were defined in \eqref{Moduli_Comp}, we have reintroduced the explicit dependece on $t^q(x)\equiv T^q|$ and
\be\label{Sugra_Laux2new}
\begin{split}
	e^{\hat K} e^{-1} {\cal L}_{\text{3-form}} 
	= & \, \frac{e^{-K}}{16} \left( {\,}^* {\cal F}_4^0 \right)^2 
	+ e^K K^{ij} {\,}^* \tilde{\cal F}_{4i} {\,}^* \tilde{\cal F}_{4j} 
	\\ 
	&+ \frac{e^{-K}}{4} K_{ij} {\,}^* {\cal F}_4^i {\,}^* {\cal F}_4^j 
	+ 4 e^K \left( {\,}^* \tilde{\cal F}_{40} \right)^2 +\call_{\rm bd}\, , 
\end{split}
\ee 
where  the four-forms $\calf_4^A$ and $\tilde\calf_{4A}$ are defined as follows
\be\label{Sugra_Fcal}
\begin{aligned}
	\mathcal F^0_4&=-F^0_4\,, \qquad \mathcal F^i_4= -F^i_4+b^iF^0_4\,,\qquad
	\tilde {\mathcal F}_{4i}= \tilde F_{4i}+k_{ijk}b^jF_4^k-\frac 12 k_{ijk}b^jb^k F^0_4\,,\\
	\tilde {\mathcal F}_{40}&= \tilde F_{40}+b^i\tilde F_{4i}+\frac 12 k_{ijk}b^ib^jF^k_4-\frac 16 k_{ijk}b^ib^jb^k F^0_4\,.
\end{aligned}
\ee 
Notice that these are identical to four-forms obtained in  \cite{Bielleman:2015ina,Carta:2016ynn} by direct dimensional reduction. The explicit form of the boundary term ${\mathcal L}_{\rm bd}$ in \eqref{Sugra_Laux2new} can be found in \cite{Farakos:2017jme}. It ensures that the variational principle for the gauge three-forms is well defined.

From the three-form Lagrangian \eqref{Sugra_Laux2new} it is clear that this dual description produces a dynamically generated potential. In fact, the integration of the equations of motion which follow from  \eqref{Sugra_bos} produces the following expressions involving  $2n+2$ integration constants $e_A$  and $m^A$ such that  
\be\label{F_40m}
\begin{aligned}
	-4e^{-\hat K}e^K\,{}^*\!\tilde {\mathcal F}_{40}=m^0\,,\qquad
	-e^{-\hat K}e^K\,K^{ij}\,{}^*\!\tilde {\mathcal F}_{4j}=m^i-m^0 b^i\equiv p^i\,,\\
	-\frac 14 e^{-( K+\hat K)}\,K_{ij}\,{}^*\!{\mathcal F}_{4}^{{j}}=e_i + k_{ijk}b^jm^k-\frac 12 k_{ijk}b^jb^k\,m^0\,\equiv \rho_i\,,\\
	-\frac 1{16}e^{-(K+\hat K)}\,{}^*\!\mathcal F^0_4=e_0+b^i e_i+\frac 12k_{ijk}b^ib^jm^k-\frac 16 k_{ijk}b^ib^jb^km^0\equiv \rho_0\,.
\end{aligned}
\ee
These correspond to the on-shell values of the four-forms obtained in \cite{Bielleman:2015ina,Carta:2016ynn} by dimensionally reducing the ten-dimensional Hodge duality relations between the type IIA R-R field-strengths \eqref{FSExp}. Substituting  \eqref{F_40m} back into the bosonic Lagrangian \eqref{Sugra_bos} and taking into account also the boundary terms, we get the following scalar potential
\be
\begin{aligned}\label{Veff}
	V = e^{\hat{K}} \Big[ 16 e^K\, \rho_0^2 + 4 e^{K}\,K^{ij}\rho_i \rho_j + e^{-K}\,K_{ij}p^ip^j+ \frac14{(m^0)^2} e^{-K} \Big].
\end{aligned}
\ee
It coincides with the type IIA R-R flux potential obtained in \cite{Louis:2002ny,Grimm:2004ua}. However, in the above description, the constants $(e_0,e_i,m^i,m^0)$ which enter the definition of  $(\rho_0,\rho_i,p^i,p^0)$, are determined by the expectation values of the four-forms $F_4^A$ and $\tilde{F}_{4A}$. 

We have thus obtained the 
manifestly supersymmetric dual formulation of effective theories describing a certain class of type IIA string compactifications.

\section{Conclusions}
In this contribution we have reviewed the non-linear duality procedure of \cite{Farakos:2017jme} which relates the usual chiral multiplets to the three-form multiplets. The core of the dualization procedure is the exchange of the coupling constants appearing in the chiral field superpotential to appropriate combinations of expectation values of real four-form field-strengths. Owing to the superspace formulation, the final output is a manifestly $\caln=1$ supersymmetric Lagrangian which includes three-form multiplets. 

Among other possible applications, this formulation provides a starting point for generalizing (in a manifestly supersymmetric framework) the Brown-Teitelboim  mechanism \cite{Brown:1987dd,Brown:1988kg} along the lines of \cite{Bousso:2000xa} and extending the results of \cite{Ovrut:1997ur,Huebscher:2009bp,Bandos:2010yy,Bandos:2012gz,Kuzenko:2017vil, Buchbinder:2017qls} on coupling the three-form supergravity-matter systems to supermembranes.

\if{}
Even though the setting of \cite{Farakos:2017jme} is quite general, here we have focused on the simple rigid Polonyi model and on the effective supergravities stemming from  compactifications of type IIA string theory in presence of R-R fluxes.  Further work is needed if one wishes to include NS-NS fluxes or D-branes. 
\fi

\section*{Acknowledgments} 
We thank  Irene Valenzuela for useful discussions. Work of F.F. is supported in part by the Interuniversity Attraction Poles Programme initiated by the Belgian Science Policy (P7/37) and the KU Leuven C1 grant ZKD1118 C16/16/005. 
Work of D.S. was supported in part by the Australian Research Council project No. DP160103633 and by the Russian Science Foundation grant 14-42-00047 in association with Lebedev Physical Institute.


\providecommand{\href}[2]{#2}\begingroup\raggedright\endgroup

\end{document}